\documentclass[sigconf]{acmart}

\usepackage{booktabs} 
\usepackage{graphicx}
\usepackage{longtable}
\usepackage{ifthen}

\providecommand{\tightlist}{%
  \setlength{\itemsep}{0pt}\setlength{\parskip}{0pt}}

\makeatletter
\let\oldlt\longtable
\let\endoldlt\endlongtable
\def\longtable{\@ifnextchar[\longtable@i \longtable@ii}
\def\longtable@i[#1]{\begin{figure*}[tbh]
\onecolumn
\begin{minipage}{1\columnwidth}
\oldlt[#1]
}
\def\longtable@ii{\begin{figure*}[tbh]
\onecolumn
\begin{minipage}{1\columnwidth}
\oldlt
}
\def\endlongtable{\endoldlt
\end{minipage}
\twocolumn
\end{figure*}}
\makeatother

\newboolean{showcomments}
\setboolean{showcomments}{true}         
\setboolean{showcomments}{false}
\ifthenelse{\boolean{showcomments}}
  {\newenvironment{changed}
    {
    \begin{center}
    \begin{tabular}{|p{1\columnwidth}|}
    \color{blue}

    }
    { 
    \end{tabular} 
    \end{center}
    }
  }
  {\newenvironment{changed}
    {  }
    {  }
  }

\setcopyright{rightsretained}

\acmDOI{10.475/123_4}

\acmISBN{123-4567-24-567/08/06}

\acmConference[SEEM 2018]{ International Workshop on Software Engineering Education for Millennials}{June 2018}{Gothenburg, Sweden}
\acmYear{2018}
\copyrightyear{2018}

\acmArticle{4}
\acmPrice{15.00}

\begin{document}
\title{Integrating Software Engineering Key Practices into an OOP Massive In-Classroom Course: an Experience Report}

\author{Marco Torchiano}
\orcid{0001-5328-368X}
\affiliation{%
  \institution{Politecnico di Torino}
  \streetaddress{C.so Duca degli Abruzzi 24}
  \city{Torino}
  \state{Italy}
  \postcode{10129}
}
\email{marco.torchiano@polito.it}

\author{Giorgio Bruno}
\affiliation{%
  \institution{Politecnico di Torino}
  \streetaddress{C.so Duca degli Abruzzi 24}
  \city{Torino}
  \state{Italy}
  \postcode{10129}
}
\email{giorgio.bruno@polito.it}


\renewcommand{\shortauthors}{M.Torchiano and G.Bruno}
\renewcommand{\shorttitle}{Integrating SE Key Practices into an OOP Massive In-Classroom Course}

\begin{abstract}
Programming and software engineering courses in computer science curricula
typically focus on both providing theoretical knowledge of programming languages and
best-practices, and developing practical development skills. In a massive
course -- several hundred students -- the teachers are not able to adequately
attend to the practical part, therefore process automation and incentives to students
must be used to drive the students in the right direction.

Our goals was to design an automated programming assignment infrastructure
capable of supporting massive courses. The infrastructure should
encourage students to apply the key software engineering (SE) practices
-- automated testing, configuration management, and Integrated Development Environment (IDE) -- 
and acquire the basic skills for using the corresponding tools.

We selected a few widely adopted development tools used to support the
key software engineering practices and mapped them to the basic
activities in our exam assignment management process.

This experience report describes the results from the past academic year.
The infrastructure we built has been used for a full academic year and
supported four exam sessions for a total of over a thousand students. 
The satisfaction level reported by the students is generally high. 



\end{abstract}

\copyrightyear{2018} 
\acmYear{2018} 
\setcopyright{acmcopyright}
\acmConference[SEEM'18]{SEEM'18:IEEE/ACM International Workshop on Software Engineering Education for Millennials }{May 27-June 3, 2018}{Gothenburg, Sweden}
\acmBooktitle{SEEM'18: SEEM'18:IEEE/ACM International Workshop on Software Engineering Education for Millennials , May 27-June 3, 2018, Gothenburg, Sweden}
\acmPrice{15.00}
\acmDOI{10.1145/3194779.3194786}
\acmISBN{978-1-4503-5750-0/18/05}

%
%



\maketitle

\section{Introduction}\label{introduction}

Software development eventually consists in delivering working code~\cite{beck2001manifesto}. Computer science and the software-related part of computer engineering should teach programming and on top of that provide software engineering skills.

At Politecnico di Torino, Italy, the first course introducing a
``\emph{modern}'' programming language is the Object Oriented
Programming (OOP) course where the language of choice is Java. For historical
reasons the BSc degree in computer engineering does not include a
Software Engineering course, therefore we decided to provide the basic
Software Engineering knowledge in the OOP course.

\begin{changed}
The students attending the course are Millennials: they were born before 1996.
While previous programming courses in the curriculum adopted paper-based exams,
we opted for a computer-based exam to leverage the technology familiarity
of ``digital natives''.
\end{changed}

The course, in addition to the Java language (version 8), provides an
introduction to UML~\cite{Rumbaugh2004}, design patterns~\cite{GoF} and basic software engineering
practices. It basically follows the indications provided in \cite{CS2013}.
The three key practices that we integrated in the course are:

\begin{itemize}
\tightlist
\item
  automated testing: represents a clear step from informally trying the
  program to a formalized and repeatable verification activity,
\item
  configuration management: introduces a standard way of versioning code
  and keeping a common shared repository,
\item
  integrated development environment: provides basic features supporting
  coding, e.g.~code completion, language specific presentation,
  automatic incremental build, error highlighting, and automatic
  code refactoring~\cite{fowler1999refactoring}.
\end{itemize}

Such practices are meant to develop software testing and software configuration skills as recommended in SWECOM 1.0~\cite{SWECOM1_0}.
Moreover, automated testing appears particularly suited to responde to  the Millennials' need for frequent feedback~\cite{Myers2010}.

A particularly tough challenge in the introduction of such practices is
represented by the size of the course: the largest of the three parallel
instances counts hundreds of newly enrolled students (330 for a.y.
2017/18). The course is taught in presence and offers practical sessions
in the lab facilities of the university. We can define it as a
Massive In-Classroom Course (MICC).
A MICC, as opposed to a MOOC, exhibits the following characteristics:

\begin{itemize}
\item the numbers are smaller than on-line courses, but still large for regular university courses; 
\item while there are videolectures, they just record the lectures held in classroom therefore they are neither primarily designed nor optimized for autonomous fruition;
\item the practical organization must enable anybody to attend in person all the educational activities (both lectures and labs);
\item it is not necessarily open, although all materials for the OOP course are freely available online.
\end{itemize}

This paper reports the experience in integrating a few key software engineering practices in
the OOP course by means of specific technologies. 
In particular we show how the devised solution was able to address both organizational and learning objectives.
On one side, such technologies support the management of assignments in the course, on the
other side they are required to perform essential tasks thus stimulating
the students to acquire the related basic skills.

First of all, section \ref{sec:context} presents the context of the course and the detailed motivation that lead us to this course implementation. Then section \ref{sec:technology} provides the details of how such key SE practices have been realized in the course. After that, section \ref{sec:discussion} discusses the educational implications, the main issues encountered, and the lessons learned.

\section{Context and Motivation}
\label{sec:context}

The Object-Oriented Programming course is located in the second year of
the Bachelor degree in Computer Engineering\footnote{Computer
  Engineering BSc. syllabus:
   \href{https://goo.gl/UMEu4y}
  {https://didattica.polito.it/ pls/portal30/gap.a\_mds.espandi2?p\_a\_acc=2018\&p\_sdu=37\& p\_cds=10\&p\_lang=EN}
}
at Politecnico di Torino. The first year encompasses fundamental topics
for all engineering disciplines (maths, computer science, physics), the
second year introduces general ICT and computer engineering topics:
circuit theory, algorithms and data structures, object-oriented
programming, and databases. The third year focuses on more advanced
topics, e.g.~operating systems, computer networks, communications,
electronics.

The OOP course introduces Object-Oriented programming using the Java
programming language and provides basic knowledge of software
engineering. The main contents are:

\begin{itemize}
\item
  Basic OO features (1 ECTS\footnote{ECTS stands for European Credit
    Transfer and Accumulation System and is a standard means for comparing
    the ``volume of learning based on the defined learning outcomes and
    their associated workload'' for higher education across the European
    Union}) including the OO paradigm, Java, classes and attributes,
  visibility, basic types, and practical skills concerning the Eclipse
  IDE.
\item
  Inheritance, interfaces, and advanced features (2 ECTS), including
  functional interfaces, lambda expressions, exceptions, and generic
  types.
\item
  Standard libraries (3 ECTS) including the Collections framework,
  streams, files, dates, threads, and GUIs.
\item
  Software Engineering principles (2 ECTS), including the Software life
  cycle, UML, Design Patterns, Configuration management, Testing. The
  latter two subtopics provide basic skills in
  Subversion\footnote{https://subversion.apache.org} and 
  JUnit\footnote{http://junit.org/junit4/}.
\end{itemize}

The course consists of over 70 hours in the classroom, including both
lectures introducing the topics and live coding sessions presenting and
discussing programming assignment solutions, and 20 hours in the lab
dedicated to the development of programming assignments. While one could argue that the hours
in the lab should be much more, this is not practically possible since
there are limited lab facilities that are shared with several other
courses, e.g.~there are 20 replicas of the basic Computer Science course with about 200 students each.

The OOP course is run in three replicas, two in Italian and one
in English, with 330, 270, and 115 enrolled students respectively.

The exam process is sketched by the activity diagram shown in
Figure 1 and encompasses a few steps:

\begin{enumerate}
\item
  the teacher prepares an initial project and uploads it;
\item
  during the exam, the students develop a small program\footnote{A few examples of the required program are available\\ at: http://softeng.polito.it/courses/02JEY/exams/} in two hours, while sitting in the lab;
\item at the end of the exam the students must submit their program;
\item meanwhile, the teacher has prepared an acceptance test suite;
\item
  after the exam, the teacher assesses the functionality of the program versus the test suite;
\item
  the students have to fix or complete the program in order to make it pass all the
  tests in the suite;
\item
  the teacher grades the work done by the students.
\end{enumerate}

The assignment consists of:

\begin{itemize}
\item a requirements document, usually made up of four or five sections that are designed to be implemented incrementally, because the features required in a section make use of the ones defined in previous sections. The requirements describe a set of classes and their methods;
\item an initial project, containing skeletal classes, i.e. classes with the methods called by the tests but with minimal bodies returning fixed values (e.g. \texttt{null}); the project can be opened with the reference IDE and is syntactically correct;
\item an example class, containing a \texttt{main()} method that exercises the most relevant methods described in the requirements. It is intended to clarify the requirements and to provide the students with a basic testing tool.
\end{itemize}

\begin{figure*}[tbp]
\centering
\includegraphics[width=0.75\linewidth]{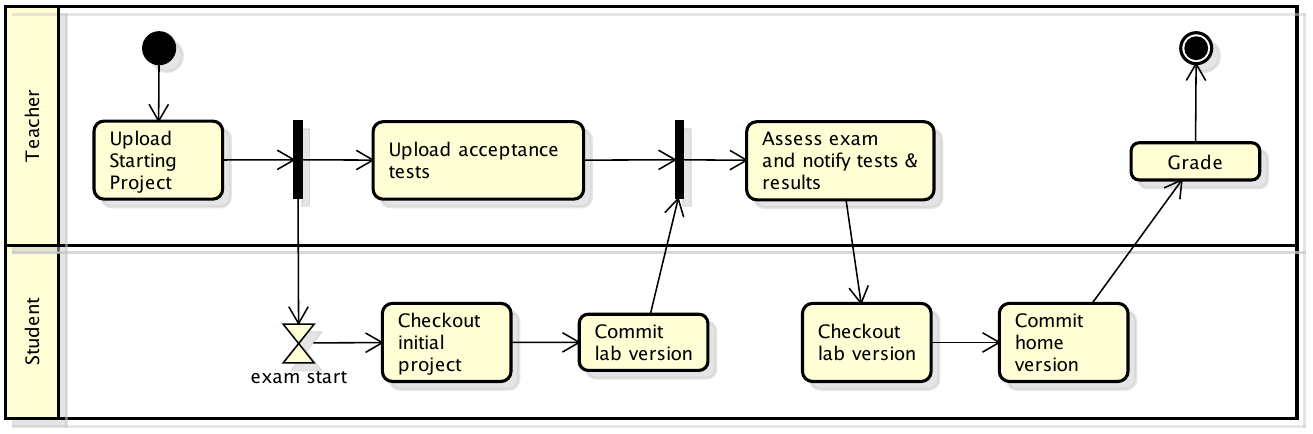}
\caption{Exam procedure}
\end{figure*}

The evaluation is computed on the basis of the functional compliance --
both in terms of correctness and completeness -- of the program.
Such an approch has been inspired by the agile manifesto principle "\textit{Working software is the primary measure of progress}"~\cite{beck2001manifesto}.

More in detail, the tests are packaged into a \texttt{.jar} file containing both class files and
source files. This is done to avoid both unintended and malicious modifications to the test suite.

In practice the grade is computed on the basis of two indicators:

\begin{itemize}
\item
  Percentage of acceptance tests passed by the lab version (\(S\)),
\item
  Code churn (\(M\)) applied to make the program pass all the tests. 
\end{itemize}

Code churn~\cite{Khoshgoftaar1996} is the amount of added and modified lines of code; it is a very simple measure of the quantity of code modification. 

The former indicator provides a coarse grained assessment of the
functional compliance from an end-user point of view, the latter
represents a fine grained evaluation and is a proxy measure of the rework needed to
fix defects (correctness) and to complete unimplemented
features (completeness).

The basic formula to compute the grade is:

\[
Grade = c_0 + c_1 \cdot \left( S + (1-S) \frac{c_2}{c_2+M}  \right)
\]

Where the constants \(c_0\), \(c_1\) and \(c_2\) are adjusted case by
case based on the difficulty of the exam.

Given the above formula:

\begin{itemize}
\item
  when a large amount of modifications is applied the grade is
  essentially defined by the percentage of tests passed
\item
  as the percentage of passed tests get lower the component inversely
  proportional to the modifications gets a higher weight.
\end{itemize}

An important aspect of this evaluation approach is that the completeness and correctness of the program delivered in the lab is evaluated by comparison. 
The reference program is the fully working version submitted from home, after the exam; it is a natural evolution carried out by the same student who wrote it initially in the lab. 
A possible alternative would be to use a predefined solution developed by the teacher as a reference, but its adequacy could be low for the following reasons:

\begin{itemize}
\tightlist
\item
  since there is no single solution for any given problem, the comparison
  with a predefined solution could penalize different -- possibly even
  better -- solutions;
\item
  the amount of work needed to complete the program and to fix defects
  can reasonably be estimated only by comparing the original one with the
  evolved version.
\end{itemize}

The approach also encourages the students to understand the requirements, identify a design and then  work on the requirements, one by one, developing fully working code (possibly just for a subset of the requirements) rather than write a complete solution in a single \textit{big-bang}, which typically does not work.

The rationale behind such an approach is that, in real-world terms, it is
better to have a program that performs correctly on a subset of
requirements that a program that is almost complete but crashes at the beginning
and eventually does nothing.

The above assessment method is fully automated and can be applied to
large numbers of delivered projects producing objective and unbiased
grades.

The current approach  presented in this paper is an evolution of the one developed originally in
2003 and described in~\cite{Torchiano2009}.

The approach was updated in response to several challenges:

\begin{itemize}
\tightlist
\item
  the number of students enrolled raised significantly in the latest
  years from around 200 per year to roughly 700 in the current academic
  year, making this course a real Massive In-Classroom Course; therefore
  the solution must be scalable and robust;

\item 
  the teaching staff is very limited: three teachers lecturing three parallel tracks (to fit lecture halls hosting 250 students at most), plus three teaching assistants supporting the students in the labs;
\item
  the lectures are video recorded and this encourages the students to attend
  the course remotely. In particular the students should be able to work
  autonomously on their assignments -- e.g.~at home -- due both to
  personal reasons and to crowded labs; therefore the assignment
  management framework must be based on tools that can be easily installed on their PCs;
\item
  the instruments and tools should enable the students to
  acquire skills directly usable in a real-world setting; therefore the tools
  should be widely adopted in practitioner communities;
\item
  the course content was extended to include basic SE practices,
  so as to encourage the students to adopt or at
  least become acquainted with basic software engineering practices, i.e.:

  \begin{itemize}
  \tightlist
  \item
    automated testing,
  \item
    configuration management,
  \item
    integrated development environment (IDE).
  \end{itemize}
\end{itemize}

To better characterize the learning outcomes, we can refer to taxonomies that
describe curricula objectives in terms of topics and levels of understanding. 
In particular, Bloom's taxonomy~\cite{bloom1956taxonomy, anderson2001taxonomy} classifies learning achievements into six different cognitive levels: knowledge, comprehension, application, analysis, synthesis and evaluation. While the educational goals in the programming part of the course clearly address all the six levels of the taxonomy, the software engineering part only addresses the lower levels of the taxonomy.


\section{Technological platform}\label{sec:technology}

The technological solution we developed is based on a few technologies
that both implement software engineering best-practices and cover a key
role in the exam and assignment management process described above.

%

The SE areas we decided to cover with technologies and assignment
related activities are:
\begin{itemize}
\item Automated Testing using JUnit
\item Configuration Management using Subversion
\item IDE as Eclipse
\end{itemize}

In addition we had to set for a robust method for authentication and
authorization to be used during exams. We decided to use SVN
authentication as the basic technology.

\subsection{Automated Testing}\label{testing}

Testing is a key technique for the Verification and Validation phase in
any software development process~\cite{runeson2006survey}, in particular automated unit testing has
gained much attention in recent years. 

JUnit is the de-facto standard for writing automated tests in Java~\cite{beck1998test}. 
While its original purpose was to write unit tests, it is
also widely used as the basis for UI testing and end-to-end tests.

In our approach, JUnit is used to evaluate the functional compliance of
assignments. The basic measure is the proportion of passed
test cases. The JUnit execution report is the standard feedback the
students receive both when they complete their lab assignments during the course and
right after the exam.

In terms of test automation, the main challenge is that we need to test a
huge number of programs. The peculiarity is that while in a regular
industrial setting we have large test suites to be executed on a single
program (or parts thereof), in our course we have a single (small) test
suite to be run against several hundred similar programs that provide different implementations of the same classes. 

A technical obstacle is that -- at least in theory -- we ought to start
a new Java Virtual Machine (JVM) for every project, load the tests and the classes making up the program, and run the tests. Unfortunately the VM startup and class
loading are very heavy tasks.

The solution we devised to achieve a reasonable scalability is to use a
hierarchy of Java class loaders as shown in Figure 2.

\begin{figure}[tbp]
\centering
\includegraphics[width=1\linewidth]{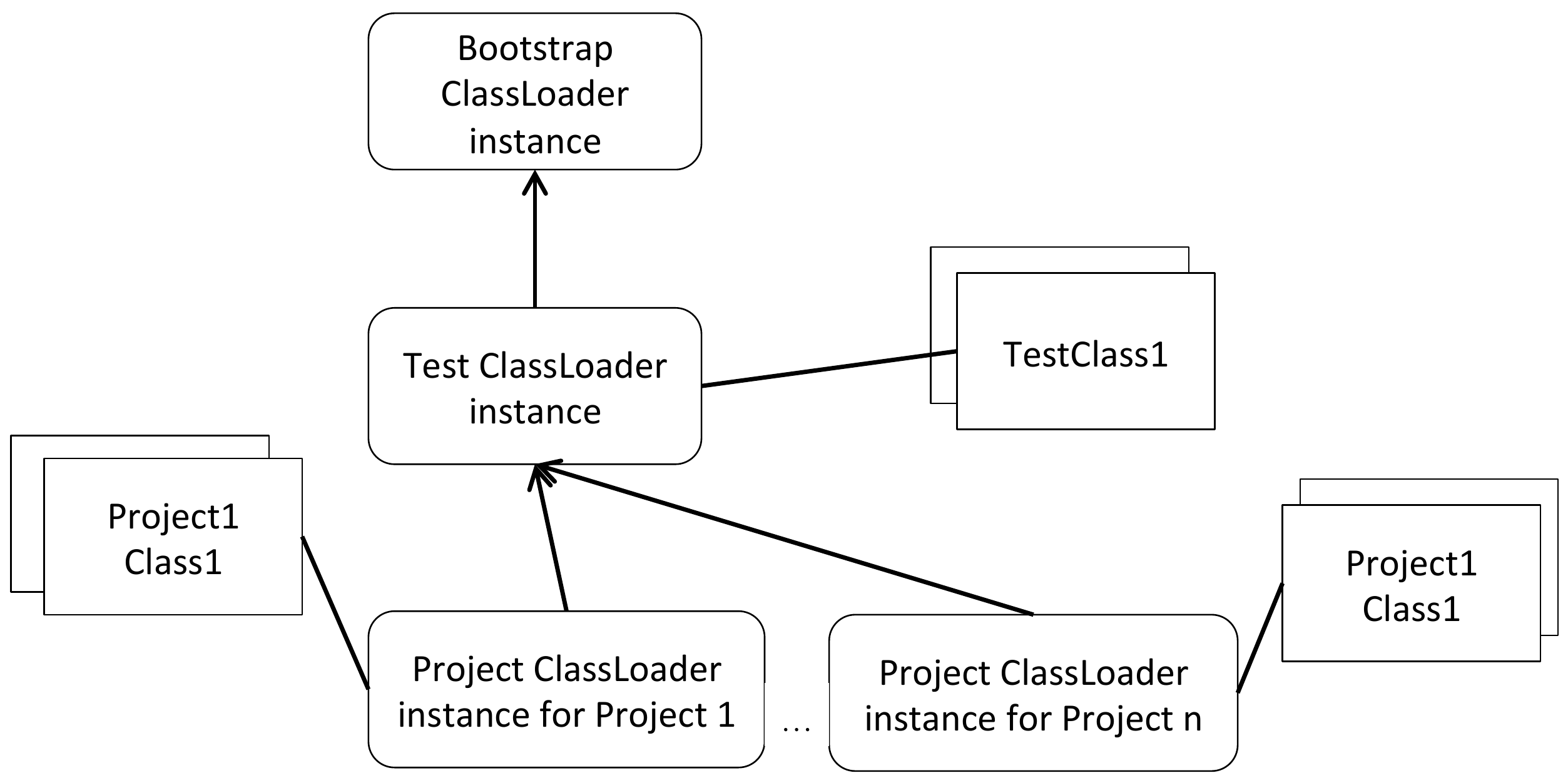}
\caption{Hierarchy of loaders used for testing.}
\end{figure}

Class loaders are classes responsible for finding and loading classes
whenever the VM needs them. A bootstrap class loader is always present
and it searches the classes in the predefined \emph{classpath}. Class
loaders are generally organized in a delegation hierarchy, so that if a
specific class loader is not able to find a class, recursively delegates
the search to its parent. In our approach a dedicated class loader class has been developed to load the test classes from a given \texttt{.jar} file. In addition we developed a project class loader that loads classes from the project path; an instance is created for each project.

The test execution starts from this more specific class loader; when a test class is needed it delegates the test class loader to get it. Since test classes are common to all the assignments, in most cases the test class loader finds them in the cache.
Such a solution has several advantages:

\begin{itemize}
\tightlist
\item
  the test classes are loaded only once by the test class loader,
\item
  any project specific class loader is isolated from the others so that
  classes with the same name can be loaded in the projects without any
  interference,
\item
  overall a single VM can be used for testing several projects.
\end{itemize}

Owing to this approach, testing projects can proceed at a rate of two projects per
second on a standard Ubuntu VM with 4 cores and 8GB RAM. Each project
typically counts five to eight classes while the test suite includes 20
test cases at least.

In terms of acquired skills and knowledge, the students are not required
to write tests -- the main reason being the short time available for the
exam, i.e.~2 hours -- though they must be able to:
%
  (i) 
  import a test suite,
  (ii)
  run the tests,
  (iii)
  understand the tests results, and
  (iv)
  identify the cause of the test failures.

In particular, for the latter ability, the students have to know what
the assert statements mean, how an expected exception is tested, and in
general they must be able to read a failure or error message, as well as to
understand a stack trace in order to locate the origin of a failure, and also to
interpret the test code to figure out the conditions that led to the failure.

The implementation of the infrastructure includes 67 Java classes for a total of 6700 LOCs.

\subsection{Configuration Management}\label{configuration-management}

Subversion (Svn) is a widespread centralized version control system~\cite{collins2004version}. 
Although its adoption has recently decreased in favor of
more modern distributed systems, such as \textit{git}~\cite{swicegood2008pragmatic}, Svn is still widely used in industry and as far as our course is concerned it is easier to use, thus less error prone, and simpler to manage. 
In our approach Subversion is used to give the assignments to the students as well as to collect their implementations, and this takes place both during the course and at the exam.

While Svn can support concurrent development with a
Copy-Modify-Merge approach, and can manage different threads of execution
using branches, the course makes use only of the basic versioning features.

In practice the assignment life cycle is supported by Svn as follows:

\begin{enumerate}
\item
  the teacher commits an initial version of a Java project
  together with an acceptance test suite to a \emph{master} repository,
\item
  the initial project is committed to all student repositories
  by the teacher using a simple script,
\item
  the students check-out the initial project and start working on it,
\item
  the students commit the results of their work to their own
  repositories,
\item
  the teacher checks out the latest version of the projects available in
  the repositories and runs the tests on them.
\end{enumerate}

The latter step is performed using the multiple classloaders approach described in the previous sub-section.


The main challenges faced in customizing Svn for the purpose of the
course were as follows:

\begin{itemize}
\tightlist
\item
  the students must have isolated personal repositories, 
  so that no interference can occur by mistake;
\item
  during the exam, the students must not be able to access other students'
  repositories, to avoid plagiarism;
\item
  during the exam, the students must be able to access their repositories as
  soon as the exam begins; therefore the repositories must be created in
  advance;
\item
  the students must not be able to keep working after the exam deadline
  has elapsed.
\end{itemize}

The isolation can be obtained by means of a single repository containing one
subfolder per student and adequate permissions. Alternatively, one
repository per student can be created with the student having access to
her own repository only. While the former is more efficient, it is less
isolated: every time a student performs a commit, the revision number is
incremented for every other students too. For this reason, even if
it is more expensive we opted for the one repository per student
solution.

During the course, a student sharing his credential with a
colleague is generally not a problem and can foster collaboration. But, such
behavior must be prevented during the exam. The solution is to create a new
repository for each student who signed up for the exam. Then each repository is
populated with a copy of the initial project and the credentials for the
repositories are handed to the students at the beginning of the exam.

The creation of an Svn repository, on our server, typically
takes 4-5 seconds, therefore the repositories must be created in
advance, at the beginning of the course and before each exam session.

During the exam, students are allowed to commit their projects as many
times as they wish. At the end of the exam, the teacher annotates the
actual end time; only commits performed before the end time are taken into account.

In addition Svn was used to make the tests available to the students on a dedicated test repository.
%
%
Sometimes errors can be found in tests after the reports have been sent to the students.
By using Subversion we can update the test \texttt{.jar} inside the repository and notify the students via email.

In terms of acquired skills and knowledge, the students must learn a few
basic tasks:

\begin{itemize}
\tightlist
\item
  performing the check-out of a project from a repository,
\item
  performing the commit of a project to a repository.
\end{itemize}

In addition, during the course the students are encouraged to perform
frequent commits when developing a project. During the
exam, they are invited to commit after implementing each requirement and
explicitly instructed that is \emph{safer} to commit 10 minutes
before the deadline. The goal of the course is to make the students
familiar with the elementary configuration management operations that
are at the basis of any workflow they will adopt in the future.

The management of the repositories has been implemented using 16 scripts
in bash and python, for a total of 910 and 827 LOCs respectively.

\subsection{Java IDE}\label{java-ide}

The usage of an IDE is often an implicit assumption when writing code.
In our course we opted for Eclipse\footnote{http://www.eclipse.org}
because historically it was one of the most widespread IDEs and due to the
fact that it is an open-source product.

The Eclipse Java IDE is the reference IDE that is taught during the
course. The configuration management and testing tasks are performed by
the students using the plug-ins for this IDE.

Eclipse is installed in all labs and the students are encouraged to
install it on their machines. We observe that while Eclipse comes with a
built-in JUnit plug-in, -- oddly enough -- it has no default built-in
plug-in for Subversion. Therefore an additional plug-in (Subversive) has
to be installed on top of the default Java IDE.


\begin{table*}[tb]
\caption{Cognitive level and specific capabilities addressed for the key
SE practices.}\label{tab:levels}
\centering
\begin{tabular}{@{}llll@{}}
\toprule
\begin{minipage}[b]{0.1\textwidth}\raggedright\strut
Level
\strut\end{minipage} &
\begin{minipage}[b]{0.25\textwidth}\raggedright\strut
Testing
\strut\end{minipage} &
\begin{minipage}[b]{0.25\textwidth}\raggedright\strut
Configuration Management
\strut\end{minipage} &
\begin{minipage}[b]{0.25\textwidth}\raggedright\strut
IDE
\strut\end{minipage}\tabularnewline
\midrule

\begin{minipage}[t]{0.1\textwidth}\raggedright\strut
Remember
\strut\end{minipage} &
\begin{minipage}[t]{0.25\textwidth}\raggedright\strut
\textbf{JUnit framework elements}
\strut\end{minipage} &
\begin{minipage}[t]{0.25\textwidth}\raggedright\strut
\textbf{Svn operation}
\strut\end{minipage} &
\begin{minipage}[t]{0.25\textwidth}\raggedright\strut
\textbf{Eclipse features}
\strut\end{minipage}\tabularnewline
\addlinespace

\begin{minipage}[t]{0.1\textwidth}\raggedright\strut
Understand
\strut\end{minipage} &
\begin{minipage}[t]{0.25\textwidth}\raggedright\strut
\textbf{Semantics of test methods and assert statements}
\strut\end{minipage} &
\begin{minipage}[t]{0.25\textwidth}\raggedright\strut
\textbf{Semantics of commands}
\strut\end{minipage} &
\begin{minipage}[t]{0.25\textwidth}\raggedright\strut
\textbf{Main tasks (e.g.~compile, run, etc.)}
\strut\end{minipage}\tabularnewline
\addlinespace

\begin{minipage}[t]{0.1\textwidth}\raggedright\strut
Apply
\strut\end{minipage} &
\begin{minipage}[t]{0.25\textwidth}\raggedright\strut
\textbf{Execute test suite}
\strut\end{minipage} &
\begin{minipage}[t]{0.25\textwidth}\raggedright\strut
\textbf{Perform \emph{check-out} and \emph{commit}}
\strut\end{minipage} &
\begin{minipage}[t]{0.25\textwidth}\raggedright\strut
\textbf{Develop and run}
\strut\end{minipage}\tabularnewline
\addlinespace

\begin{minipage}[t]{0.1\textwidth}\raggedright\strut
Analyze
\strut\end{minipage} &
\begin{minipage}[t]{0.25\textwidth}\raggedright\strut
\textbf{Understand test results}
\strut\end{minipage} &
\begin{minipage}[t]{0.25\textwidth}\raggedright\strut
\emph{Understand outcome of operations}
\strut\end{minipage} &
\begin{minipage}[t]{0.25\textwidth}\raggedright\strut
\textbf{Understand error messages}
\strut\end{minipage}\tabularnewline
\addlinespace

\begin{minipage}[t]{0.1\textwidth}\raggedright\strut
Evaluate
\strut\end{minipage} &
\begin{minipage}[t]{0.25\textwidth}\raggedright\strut
\textbf{Identify failure causes}
\strut\end{minipage} &
\begin{minipage}[t]{0.25\textwidth}\raggedright\strut
Identify conflict causes
\strut\end{minipage} &
\begin{minipage}[t]{0.3\textwidth}\raggedright\strut
\textbf{Identify defects or problems}
\strut\end{minipage}\tabularnewline
\addlinespace

\begin{minipage}[t]{0.1\textwidth}\raggedright\strut
Create
\strut\end{minipage} &
\begin{minipage}[t]{0.25\textwidth}\raggedright\strut
Write tests
\strut\end{minipage} &
\begin{minipage}[t]{0.25\textwidth}\raggedright\strut
Merge conflicts
\strut\end{minipage} &
\begin{minipage}[t]{0.25\textwidth}\raggedright\strut
Set-up a project
\strut\end{minipage}\tabularnewline
\bottomrule
\end{tabular}
\end{table*}

\section{Discussion}\label{sec:discussion}

\subsection{Learning objectives}\label{learning-objectives}

Table \ref{tab:levels} reports the six taxonomy levels and the corresponding
capabilities addressed with respect to the three key SE areas included
in our course. In the table, the capabilities addressed by the
course are shown in \textbf{bold}, those partially addressed in
\emph{italic}, and the others, not addressed, in a regular
font.

The cognitive levels addressed are first needed in the lab assignments the
students have to perform during the course and then they are required in the exam. Therefore we are confident that the students passing
the exam achieved those levels to a good degree of completeness.

The \emph{Analyze} level for configuration management is only partly
addressed because no concurrent development is used in the course,
therefore no conflict will take place: this is an activity students
learn in lectures but never experience in practice. For this reason, the \emph{Evaluate} level is not addressed either.

The \emph{Create} level is not addressed for any of the three key areas.
As to testing, writing tests is a time consuming activity that
cannot fit in the tight schedule (2 hours) allowed for the exam. As far
as configuration management is concerned, the lack of concurrent development makes it impossible to
apply merge operations. Regarding the Java IDE, all assignments start with students importing
pre-defined Eclipse projects from Svn, therefore the project set-up phase is not put into practice.

Concerning the basic skills we observed one important point: even though most students are able to perform correct Svn operations, they tend to apply a minimalistic workflow. Students are encouraged to perform a commit after completing each requirement section, nevertheless most of them tend to perform fewer commits, just the barely minimum to abide by the exam rules: a commit at the end of the exam session and a commit after the session.

Given an assignment whose requirements contain $r$ sections, the recommended process entails at least $r+1$ commits: one for each requirement section plus one from home after the exam. This is a very simple, though approximate, criterion to identify compliant students.

We analyzed the number of commits performed by the students on their exam repositories.
Out of 1008 repositories -- corresponding to the bookings --
we found that 25\% of them contained only the initially project and no student commit. These are untouched projects: students that either booked the exam but did not show up or decided to quit during the exam.

\begin{figure*}[tb]
\includegraphics[width=0.95\textwidth]{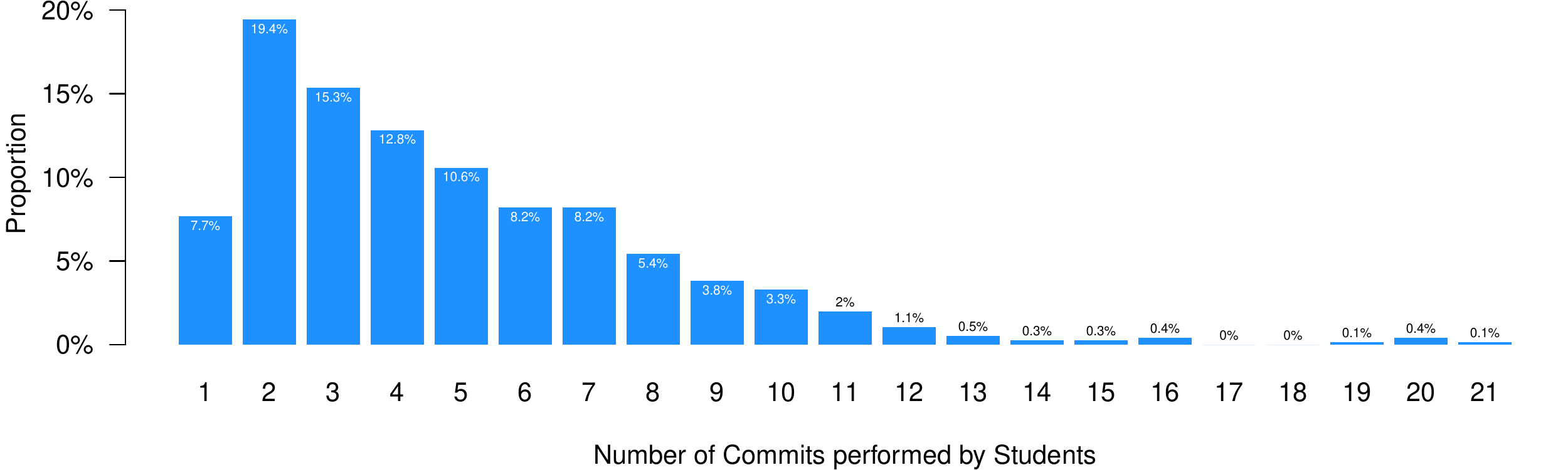}
\caption{Distribution of student commits.}
\label{fig:commits}
\end{figure*}

Excluding the untouched projects, the distribution of the number of commits for the students who actually attended the exams is shown in Figure \ref{fig:commits}.

There is a small percentage of students (7.7\%) who performed just 1 commit, i.e. they committed a version in the lab during the exam but did not completed their programs at home; they are the exam dropouts. A larger share of the students (92.3\%) performed at least two commits, i.e. one in the lab and one from home.

Table \ref{tab:compliant} reports, for each exam session, the number of touched repositories and the proportion of dropouts and compliant students. We observe that overall 39\% of students complied with the recommended process. The first two exam sessions -- closely following the end of the course -- exhibit a higher compliance, 44\% and 41\% respectively.

\begin{table}[tb]
\caption{Process compliant students}
\label{tab:compliant}
\begin{tabular}{lrrr}
\toprule
\textbf{Session} & \textbf{Students} & \textbf{Dropout} & \textbf{Compliant} \\
\midrule
June 2017 &  334 & 7.5\% & 44.3\% \\
July 2017 &  258 & 6.6\% & 41.5\% \\
Sept 2017 &  101 & 15.8\% & 22.8\% \\
Jan 2018  &   63 & 0.0\% & 23.8\% \\
\addlinespace
\textit{All} & 756 & 7.7\% & 38.8\% \\
\bottomrule
\end{tabular}
\end{table}

\subsection{Issues}\label{issues}

The first instance of the course using the infrastructure described
above was given in a.y. 2016/17. The set-up was used both during the
course (from March to June 2017) and for the exam sessions. We managed
four exam sessions (June, July and September 2017, and January 2018) for a
total of 629 exams graded. Given the huge number of students we
encountered several problems.

We summarize here the main issues that emerged during and after the
exam sessions: 

\begin{itemize}
\tightlist
\item
  Several students after checking-out realized that Eclipse did not
  provide editing support (e.g.~code completion). This is typically
  due to the fact they did check-out the whole repository and not just
  the folder containing the Eclipse project; as a consequence, Eclipse is
  not able to recognize the folder as a Java project and thus cannot provide
  all Java-related supporting features.
\item
  After the exam some students got a test report showing many failures they
  could not find in their projects. The cause for this lies in a late
  commit, i.e. a commit performed after the exam deadline.
\item
  After the exam, some students got no test reports because the
  projects submitted contained errors that prevented a successful compilation.
  Despite the invitation issued 10 minutes before the end of the
  exam, several students continued to work on the code rather than
  checking the code for errors.
\item
  Sometimes students get compilation errors they were not able to see in
  the lab within their IDE. In our experience this is due to a few
  causes:

  \begin{itemize}
  \tightlist
  \item
    the Eclipse uses its own (incremental) compiler that in a few cases
    -- e.g.~type inference for generic types -- behaves differently from
    the Oracle JDK compiler we use to compile the project before testing;
  \item
    the Eclipse IDE, when suggesting imports in case of undefined
    classes or interfaces, usually provides a list of all compatible
    elements, e.g.~for \texttt{Collections} it includes of course the
    \texttt{java.util.Collections} class as well as, e.g.
    \texttt{com.sun.xml.internal.ws.policy.\ privateutil.PolicyUtils.Collections}.
    The latter class is usually not present in a clean JDK installation
    and the corresponding \texttt{import} is marked as an error during
    compilation.
  \end{itemize}
\item
  During the first exam session, the students in a lab were not able to
  connect to the Svn repository. This problem was caused by a
  misconfiguration in the web proxy and firewall in just one lab.
\end{itemize}

As a side note, we also encountered some weird issues unrelated to the
topics covered in the course. A few students in every exam session
typically call for help because suddenly the editor in Eclipse is
overwriting their code instead of inserting new characters: this is due
to the fact that the students inadvertently pushed the \emph{Ins} button
on the keyboard thus switching from insert to overwrite mode. We speculate
that such \emph{magic Ins key} problem is due to some students being
used to small factor laptop keyboards that do not have a dedicated
\emph{Ins} key.

Another issue that emerged while discussing with colleages is the suitability of the Eclipse IDE. In the last year the Eclipse market share\footnote{http://www.baeldung.com/java-in-2017} (40\%) appears to be shrinking in favor of IntelliJ IDEA (46\%), which according to colleagues provide a more modern and usable environment.

\subsection{Lessons learned}\label{lessons-learned}

We collected a number of critical issues that we intend to overcome in the next version of the course.

  \textbf{Students are not able to use the basic tools}: this is
  particularly true for Subversion (as reported above) but sometimes it
  happens they are not familiar with Eclipse or even with the PCs
  available in the university lab. The lesson we learned is that the countermeasure
  is to force or provide incentives for the students to get familiar with
  the tools \emph{before} sustaining the exam. Currently the assignments
  proposed to the students during the course are not mandatory. A
  possible mitigation to this problem may be to give additional
  points in the final grade if the students complete a specific assignment that
  requires basic skills (e.g.~Subversion).

  \textbf{The development environment might differ in part from the
  testing environment}: this is typically due to the compiler (Eclipse
  own compiler vs.~JDK javac), the \emph{classpath} (Eclipse Java project
  vs.~clean JDK), or the operating systems (Windows in the lab
  vs.~Ubuntu for the test server). The consequences of this issue can be
  significantly reduced by implementing a simplified \emph{Continuous
  Integration}~\cite{duvall2007continuous} infrastructure. Every commit goes through compilation and
  testing and the results are reported back to the students. Such a
  feedback would enable the students to understand what the problem is
  in the testing environment.

  \textbf{Students assume they can work in a new environment just
  because they used a similar one}: several students -- because of the
  crowded lecture rooms and labs, the availability of video recorded
  lectures, and the possibility of performing assignments on their own
  PCs -- tend not to attend all lectures and labs. 
  As a consequence, the day of the exam turns out to be the first time they use the lab equipment.
  \begin{changed}
  The (presumed) tech savyness and confidence of Millennials apparently
  bring them to overestimate their knowledge.
  \end{changed}

  However, forcing the students to work on their assignments in
  the lab, would restrict their freedom, and possibly
  overload both the facility and the teaching assistants. It is
  important to make sure the environment the students re-create on their
  machines is as close as possible to the lab environment. This can be
  achieved by defining very well the reference environment -- IDE and
  JDK version -- as well as ensuring the latest version -- the one that 
  the students will download most likely -- is installed in the
  lab too.
  
  \textbf{Scalable and reliable automation requires a lot of effort for
  the infrastructure}: even if the three cornerstone technologies are
  quite sound and mature, their usage in the course is peculiar and
  requires dedicated workflows to be designed as well as a suitable 
  infrastructure to be developed. For this course, during several years, over
  10KLOC of code were written mostly in Java but also in Python, Bash
  shell and Html. The recommendation is to use existing tools as far as
  possible but also to be prepared for a large effort in infrastructure
  development.

  \textbf{Whenever you rely on a server, never underestimate the
  network}: we performed tests in two (out of six) labs used for the
  exam, but not in the one that turned out to have the issue. The
  recommendation is of course that extensive testing must be performed
  in the field.

\section{Conclusions}\label{conclusions}

This paper presented a report on the experience in integrating three key
Software Engineering practices -- automated testing, configuration
management, and integrated development environment -- into a large OOP course. 
The key practices play a twofold role: first, they are instrumental to
achieve a set of educational goals, second, they are the cornerstones of
the infrastructure supporting assignment management both during the
course and at the exams.

\begin{changed}
The resources required to run the course consist in a linux server hosting the Subversion repositories, the scripts, and runnning the test correction procedure. In addition
labs large enough are required with PCs hosting the Eclipse IDE. All the required software is open-source and the additional custom software can be provided upon request.
\end{changed}

The course, as reported, has been run once in a.y. 2016/17, although it builds on almost 15 years of experience. The anonymous student satisfaction questionnaires resulted in 90\% students being
overall satisfied for the educational part. The global satisfaction
level -- also including  the logistics -- is at 83\%, mainly due to the crowded classes and labs.

For the next edition of the course we plan to put into practice the
lessons learned, the most important being the introduction of a
light-weight continuous integration feature.

Moreover for future editions we will have to consider a possible evolution of the adopted technologies (e.g. IDE and configuration mangement), taking into account both the ease of use and the popularity.

\bibliographystyle{ACM-Reference-Format}
\bibliography{refs.bib}


\begin{thebibliography}{16}


\ifx \showCODEN    \undefined \def \showCODEN     #1{\unskip}     \fi
\ifx \showDOI      \undefined \def \showDOI       #1{#1}\fi
\ifx \showISBNx    \undefined \def \showISBNx     #1{\unskip}     \fi
\ifx \showISBNxiii \undefined \def \showISBNxiii  #1{\unskip}     \fi
\ifx \showISSN     \undefined \def \showISSN      #1{\unskip}     \fi
\ifx \showLCCN     \undefined \def \showLCCN      #1{\unskip}     \fi
\ifx \shownote     \undefined \def \shownote      #1{#1}          \fi
\ifx \showarticletitle \undefined \def \showarticletitle #1{#1}   \fi
\ifx \showURL      \undefined \def \showURL       {\relax}        \fi
\providecommand\bibfield[2]{#2}
\providecommand\bibinfo[2]{#2}
\providecommand\natexlab[1]{#1}
\providecommand\showeprint[2][]{arXiv:#2}

\bibitem[\protect\citeauthoryear{{ACM/IEEE-CS Joint Task Force on Computing
  Curricula}}{{ACM/IEEE-CS Joint Task Force on Computing Curricula}}{2013}]%
        {CS2013}
\bibfield{author}{\bibinfo{person}{{ACM/IEEE-CS Joint Task Force on Computing
  Curricula}}.} \bibinfo{year}{2013}\natexlab{}.
\newblock \bibinfo{booktitle}{\emph{Computer Science Curricula 2013}}.
\newblock \bibinfo{type}{{T}echnical {R}eport}. \bibinfo{institution}{ACM Press
  and IEEE Computer Society Press}.
\newblock


\bibitem[\protect\citeauthoryear{Anderson, Krathwohl, Airasian, Cruikshank,
  Mayer, Pintrich, Raths, and Wittrock}{Anderson et~al\mbox{.}}{2001}]%
        {anderson2001taxonomy}
\bibfield{author}{\bibinfo{person}{Lorin~W. Anderson},
  \bibinfo{person}{David~R. Krathwohl}, \bibinfo{person}{P. Airasian},
  \bibinfo{person}{K. Cruikshank}, \bibinfo{person}{R. Mayer},
  \bibinfo{person}{P. Pintrich}, \bibinfo{person}{James Raths}, {and}
  \bibinfo{person}{M. Wittrock}.} \bibinfo{year}{2001}\natexlab{}.
\newblock \bibinfo{booktitle}{\emph{A taxonomy for learning, teaching and
  assessing: A revision of Bloom's taxonomy}}.
\newblock \bibinfo{publisher}{Longman}.
\newblock


\bibitem[\protect\citeauthoryear{Beck, Beedle, Van~Bennekum, Cockburn,
  Cunningham, Fowler, Grenning, Highsmith, Hunt, Jeffries, et~al\mbox{.}}{Beck
  et~al\mbox{.}}{2001}]%
        {beck2001manifesto}
\bibfield{author}{\bibinfo{person}{Kent Beck}, \bibinfo{person}{Mike Beedle},
  \bibinfo{person}{Arie Van~Bennekum}, \bibinfo{person}{Alistair Cockburn},
  \bibinfo{person}{Ward Cunningham}, \bibinfo{person}{Martin Fowler},
  \bibinfo{person}{James Grenning}, \bibinfo{person}{Jim Highsmith},
  \bibinfo{person}{Andrew Hunt}, \bibinfo{person}{Ron Jeffries},
  {et~al\mbox{.}}} \bibinfo{year}{2001}\natexlab{}.
\newblock \showarticletitle{Manifesto for agile software development}.
\newblock  (\bibinfo{year}{2001}).
\newblock


\bibitem[\protect\citeauthoryear{Beck and Gamma}{Beck and Gamma}{1998}]%
        {beck1998test}
\bibfield{author}{\bibinfo{person}{Kent Beck} {and} \bibinfo{person}{Erich
  Gamma}.} \bibinfo{year}{1998}\natexlab{}.
\newblock \showarticletitle{Test infected: Programmers love writing tests}.
\newblock \bibinfo{journal}{\emph{Java Report}} \bibinfo{volume}{3},
  \bibinfo{number}{7} (\bibinfo{year}{1998}), \bibinfo{pages}{37--50}.
\newblock


\bibitem[\protect\citeauthoryear{Bloom et~al\mbox{.}}{Bloom
  et~al\mbox{.}}{1956}]%
        {bloom1956taxonomy}
\bibfield{author}{\bibinfo{person}{Benjamin~S Bloom} {et~al\mbox{.}}}
  \bibinfo{year}{1956}\natexlab{}.
\newblock \showarticletitle{Taxonomy of educational objectives. Vol. 1:
  Cognitive domain}.
\newblock \bibinfo{journal}{\emph{New York: McKay}} (\bibinfo{year}{1956}),
  \bibinfo{pages}{20--24}.
\newblock


\bibitem[\protect\citeauthoryear{Collins-Sussman, Fitzpatrick, and
  Pilato}{Collins-Sussman et~al\mbox{.}}{2004}]%
        {collins2004version}
\bibfield{author}{\bibinfo{person}{Ben Collins-Sussman}, \bibinfo{person}{Brian
  Fitzpatrick}, {and} \bibinfo{person}{Michael Pilato}.}
  \bibinfo{year}{2004}\natexlab{}.
\newblock \bibinfo{booktitle}{\emph{Version control with subversion}}.
\newblock \bibinfo{publisher}{" O'Reilly Media, Inc."}.
\newblock


\bibitem[\protect\citeauthoryear{Duvall, Matyas, and Glover}{Duvall
  et~al\mbox{.}}{2007}]%
        {duvall2007continuous}
\bibfield{author}{\bibinfo{person}{Paul~M Duvall}, \bibinfo{person}{Steve
  Matyas}, {and} \bibinfo{person}{Andrew Glover}.}
  \bibinfo{year}{2007}\natexlab{}.
\newblock \bibinfo{booktitle}{\emph{Continuous integration: improving software
  quality and reducing risk}}.
\newblock \bibinfo{publisher}{Pearson Education}.
\newblock


\bibitem[\protect\citeauthoryear{Fowler and Beck}{Fowler and Beck}{1999}]%
        {fowler1999refactoring}
\bibfield{author}{\bibinfo{person}{Martin Fowler} {and} \bibinfo{person}{Kent
  Beck}.} \bibinfo{year}{1999}\natexlab{}.
\newblock \bibinfo{booktitle}{\emph{Refactoring: improving the design of
  existing code}}.
\newblock \bibinfo{publisher}{Addison-Wesley Professional}.
\newblock


\bibitem[\protect\citeauthoryear{Gamma, Helm, Johnson, and Vlissides}{Gamma
  et~al\mbox{.}}{1995}]%
        {GoF}
\bibfield{author}{\bibinfo{person}{Erich Gamma}, \bibinfo{person}{Richard
  Helm}, \bibinfo{person}{Ralph Johnson}, {and} \bibinfo{person}{John
  Vlissides}.} \bibinfo{year}{1995}\natexlab{}.
\newblock \bibinfo{booktitle}{\emph{Design Patterns: Elements of Reusable
  Object-oriented Software}}.
\newblock \bibinfo{publisher}{Addison-Wesley Longman Publishing Co., Inc.},
  \bibinfo{address}{Boston, MA, USA}.
\newblock
\showISBNx{0-201-63361-2}


\bibitem[\protect\citeauthoryear{{IEEE-CS}}{{IEEE-CS}}{2014}]%
        {SWECOM1_0}
\bibfield{author}{\bibinfo{person}{{IEEE-CS}}.}
  \bibinfo{year}{2014}\natexlab{}.
\newblock \bibinfo{booktitle}{\emph{Software Engineering Competency Model}}.
\newblock \bibinfo{type}{{T}echnical {R}eport}. \bibinfo{institution}{IEEE
  Computer Society Press}.
\newblock
\showISBNx{0-7695-5373-7}


\bibitem[\protect\citeauthoryear{Khoshgoftaar, Allen, Goel, Nandi, and
  McMullan}{Khoshgoftaar et~al\mbox{.}}{1996}]%
        {Khoshgoftaar1996}
\bibfield{author}{\bibinfo{person}{T.~M. Khoshgoftaar}, \bibinfo{person}{E.B.
  Allen}, \bibinfo{person}{N. Goel}, \bibinfo{person}{A. Nandi}, {and}
  \bibinfo{person}{J. McMullan}.} \bibinfo{year}{1996}\natexlab{}.
\newblock \showarticletitle{Detection of Software Modules with high Debug Code
  Churn in a very large Legacy System}. In
  \bibinfo{booktitle}{\emph{Proceedings of International Symposium on Software
  Reliability Engineering}}. \bibinfo{pages}{364--371}.
\newblock


\bibitem[\protect\citeauthoryear{Myers and Sadaghiani}{Myers and
  Sadaghiani}{2010}]%
        {Myers2010}
\bibfield{author}{\bibinfo{person}{Karen~K. Myers} {and}
  \bibinfo{person}{Kamyab Sadaghiani}.} \bibinfo{year}{2010}\natexlab{}.
\newblock \showarticletitle{Millennials in the Workplace: A Communication
  Perspective on Millennials' Organizational Relationships and Performance}.
\newblock \bibinfo{journal}{\emph{Journal of Business and Psychology}}
  \bibinfo{volume}{25}, \bibinfo{number}{2} (\bibinfo{date}{01 Jun}
  \bibinfo{year}{2010}), \bibinfo{pages}{225--238}.
\newblock
\showISSN{1573-353X}
\urldef\tempurl%
\url{https://doi.org/10.1007/s10869-010-9172-7}
\showDOI{\tempurl}


\bibitem[\protect\citeauthoryear{Rumbaugh, Jacobson, and Booch}{Rumbaugh
  et~al\mbox{.}}{2004}]%
        {Rumbaugh2004}
\bibfield{author}{\bibinfo{person}{James Rumbaugh}, \bibinfo{person}{Ivar
  Jacobson}, {and} \bibinfo{person}{Grady Booch}.}
  \bibinfo{year}{2004}\natexlab{}.
\newblock \bibinfo{booktitle}{\emph{Unified Modeling Language Reference Manual,
  The (2nd Edition)}}.
\newblock \bibinfo{publisher}{Pearson Higher Education}.
\newblock
\showISBNx{0321245628}


\bibitem[\protect\citeauthoryear{Runeson}{Runeson}{2006}]%
        {runeson2006survey}
\bibfield{author}{\bibinfo{person}{Per Runeson}.}
  \bibinfo{year}{2006}\natexlab{}.
\newblock \showarticletitle{A survey of unit testing practices}.
\newblock \bibinfo{journal}{\emph{IEEE software}} \bibinfo{volume}{23},
  \bibinfo{number}{4} (\bibinfo{year}{2006}), \bibinfo{pages}{22--29}.
\newblock


\bibitem[\protect\citeauthoryear{Swicegood}{Swicegood}{2008}]%
        {swicegood2008pragmatic}
\bibfield{author}{\bibinfo{person}{Travis Swicegood}.}
  \bibinfo{year}{2008}\natexlab{}.
\newblock \bibinfo{booktitle}{\emph{Pragmatic version control using Git}}.
\newblock \bibinfo{publisher}{Pragmatic Bookshelf}.
\newblock


\bibitem[\protect\citeauthoryear{Torchiano and Morisio}{Torchiano and
  Morisio}{2009}]%
        {Torchiano2009}
\bibfield{author}{\bibinfo{person}{Marco Torchiano} {and}
  \bibinfo{person}{Maurizio Morisio}.} \bibinfo{year}{2009}\natexlab{}.
\newblock \showarticletitle{A Fully Automatic Approach to the Assessment of
  Programming Assignments}.
\newblock \bibinfo{journal}{\emph{International Journal of Engineering
  Education}}  \bibinfo{volume}{24 (4)} (\bibinfo{year}{2009}),
  \bibinfo{pages}{814--829}.
\newblock


\end{thebibliography}

%
%
%

\end{document}